\documentclass[seceq]{ptptex}

\usepackage{graphicx}
\usepackage{wrapft}

\title{
On the extensivity of the entropy $S_q$, the $q$-generalized central limit theorem and the $q$-triplet%
}

\author{
Constantino \textsc{Tsallis} \footnote{E-mails: tsallis@santafe.edu, tsallis@cbpf.br} %
}
\inst{
Santa Fe Institute, 1399 Hyde Park Road, Santa Fe, New Mexico 87501, USA
\\
and\\
Centro Brasileiro de Pesquisas Fisicas \\ 
Rua Xavier Sigaud 150, 22290-180 Rio de Janeiro - RJ, Brazil
}

\abst{
First, we briefly review the conditions under which the entropy $S_q$ can be extensive (Tsallis, Gell-Mann and Sato, Proc. Natl. Acad. Sc. USA (2005), in press; cond-mat/0502274), as well as  the possible $q$-generalization of the central limit theorem (Moyano, Tsallis and Gell-Mann, cond-mat/0509229). Then, we address the $q$-triplet recently determined in the solar wind (Burlaga and Vinas, Physica A {\bf 356}, 375 (2005)) and its possible relation with the space-dimension $d$ and with
the range of the interactions  (characterized by $\alpha$, the attractive potential energy being assumed to decay as $r^{-\alpha}$ at long distances $r$).
}

\begin{document}

\maketitle

\section{On how can $S_q$ be extensive}
Clausius introduced in 1865 the concept of entropy, $S$, without any reference to the microscopic world, i.e., to the possible microscopic configurations of the system. Indeed, at that time, most physicists did not believe in the existence of atoms and molecules. Boltzmann introduced, a few years later, the first connection of entropy to the microscopic world, namely
\begin{equation}
S_{BG} =-k \sum_{i=1}^W p_i \ln p_i \,.
\end{equation}
This expression is compatible with everything that Clausius expected for the entropy on thermodynamical grounds, in particular to be {\it extensive}. Indeed, if a system is composed of $N$ identical and distinguishable elements (or subsystems) that are probabilistically {\it independent}, i.e., $\pi_{i_1,i_2,...,i_N}^{A_1+A_2+...+A_N}=\pi_{i_1}^{A_1}\pi_{i_2}^{A_2}...\pi_{i_N}^{A_N}$, we immediately verify that $S_{BG}(N)=NS_{BG}(1)$. Moreover, if the elements systems are {\it quasi-independent} (in a sense that needs of course more precise qualification, but which we do not address at this point), we have that $S_{BG}$ still is extensive, i.e., $S_{BG}(N) \propto N$ if $N>>1$. 

An interesting question arrives if the elements of the system are {\it far from independence}. Such is the case, for instance, when the elements have nontrivial (strictly or asymptotically)  {\it scale-invariant} correlations, i.e., when the set of probabilities associated with $N$ elements is strongly correlated to the set of probabilities associated with $(N-1)$ elements. In such a case, which appears in fact quite frequently in natural and artificial systems, it can be shown that $S_{BG}$ is {\it not} extensive. Instead, the entropy
\begin{equation}
S_q =k \frac{1-\sum_{i=1}^W p_i^q}{q-1}\;\;\;\;(q \in {\cal R}; \, S_1=S_{BG}) \,.
\end{equation}
{\it can be extensive} for a special value of the index $q$ different from unity, as shown in \cite{tsallisgellmannsato}.

The most general binary case is given by $\{ \pi_{1}^{A_1},\pi_{2}^{A_1} \}=\{ \pi_{1}^{A_1},1-\pi_{1}^{A_1} \}$ if $N=1$, by Table I if $N=2$, by Table II if $N=3$, and so on. For arbitrary $N$, it can be thought as a $N$-dimensional ``hypercube" having $W=2^N$ probabilities $\{\pi_{i_1, i_2,...,i_N}^{A_1+A_2+...+A_N}  \}$ such that $\sum_{i_1=1}^2 \sum_{i_2=1}^2 ... \sum_{i_N=1}^2 \pi_{i_1, i_2,...,i_N}^{A_1+A_2+...+A_N}=1$.

\begin{table}[htbp]
\begin{center}
\begin{tabular}{c||c|c||cccccc}
$_{A_1}\setminus^{A_2}$ &1                                                                       &2                                                                                 \\
[1mm] \hline\hline
1                                        &$\pi_{11}^{A_1+A_2}$                                     &$\pi_{12}^{A_1+A_2}$                                   &$\pi_{11}^{A_1+A_2} + \pi_{12}^{A_1+A_2}$ \\
[3mm] \hline
2                                        &$\pi_{21}^{A_1+A_2}$                                     &$\pi_{22}^{A_1+A_2}$                                   &$\pi_{21}^{A_1+A_2} + \pi_{22}^{A_1+A_2}$ \\
[3mm] \hline \hline
                                          &$\pi_{11}^{A_1+A_2} +\pi_{21}^{A_1+A_2}$   &$\pi_{12}^{A_1+A_2}+\pi_{22}^{A_1+A_2}$  &1
\end{tabular}
\end{center}
\begin{tabular}{cccccccccccc}
& &&&&     \\[1mm] 
&&&&&&    \\[3mm] 
&&&&&&    \\[3mm] 
&&&&&& 
\end{tabular}
\vspace{-2.0cm}
\caption{ {\it Joint} and {\it marginal} probabilities for $N=2$, i.e., two binary subsystems $A_1$ and $A_2$. 
}
\end{table}
 
\begin{table}[htbp]
\begin{center}
\begin{tabular}{c||c|c||}
 $_{A_1}\setminus^{A_2}$    &  1                                                                & 2                                                    \\
[1mm] \hline\hline
1                              &$\pi^{A_1+A_2+A_3}_{111}$                                        &$\pi^{A_1+A_2+A_3}_{121}$    \\   
                                &$[\pi^{A_1+A_2+A_3}_{112}]$                                      &$[\pi^{A_1+A_2+A_3}_{122}]$            \\                                               
[3mm] \hline
2                              &$\pi^{A_1+A_2+A_3}_{211}$                                        &$\pi^{A_1+A_2+A_3}_{221}$             \\
                                &$[\pi^{A_1+A_2+A_3}_{212}]$                                      & $[\pi^{A_1+A_2+A_3}_{222}]$          \\                                         
[3mm] \hline \hline
\end{tabular}
\end{center}
\vspace{0.7cm}
\caption{Joint probabilities $\pi^{A_1+A_2+A_3}_{i_1\,i_2\,i_3}$ for $N=3$. The quantities without (within) square-brackets $[\;]$ correspond to state 1 (state 2) of subsystem $A_3$. 
}
\end{table}

In general, for $N=2$, we have $\pi_{12}^{A_1+A_2} \ne \pi_{21}^{A_1+A_2}$; for $N=3$, we have $\pi^{A_1+A_2+A_3}_{112} \ne \pi^{A_1+A_2+A_3}_{121} \ne \pi^{A_1+A_2+A_3}_{211}$ and $\pi^{A_1+A_2+A_3}_{122} \ne \pi^{A_1+A_2+A_3}_{212} \ne \pi^{A_1+A_2+A_3}_{221}$; and so on for $N=4,5,...$. If we interpret $N$ as the ``time" of a time series, it is precisely these discrepancies which characterize ``long-memory". We shall consider from now on the simple case where there is no ``long-range" memory, i.e.,  $\pi_{12}^{A_1+A_2} = \pi_{21}^{A_1+A_2} \equiv r_{21}$ (with $\pi_{11}^{A_1+A_2} \equiv r_{20}$ and $\pi_{22}^{A_1+A_2} \equiv r_{22}$), 
$\pi^{A_1+A_2+A_3}_{112} = \pi^{A_1+A_2+A_3}_{121} = \pi^{A_1+A_2+A_3}_{211} \equiv r_{31}$ and  $\pi^{A_1+A_2+A_3}_{122} = \pi^{A_1+A_2+A_3}_{212} = \pi^{A_1+A_2+A_3}_{221} \equiv r_{32}$ (with $\pi^{A_1+A_2+A_3}_{111} \equiv r_{30}$ and $\pi^{A_1+A_2+A_3}_{222} \equiv r_{33}$), and so on. This simple case can be represented on a triangle, as shown in Table III.

We may impose on this triangle a very special correlation, namely the {\it Leibnitz rule} (see details in \cite{tsallisgellmannsato}), i.e., $r_{N,n}+r_{N,n+1} =r_{N-1,n}\,\;\;\;(n=0,1,...,N-1;\,N=2,3, ...)$. This correlation makes the system scale-invariant in the sense that all the marginal probabilities associated with the $N$-system coincide with the joint probabilities of the $(N-1)$-system. The system may satisfy Leibnitz rule either strictly or only asymptotically when $N \to\infty$. We will say respectively that the system is {\it strictly} or {\it asymptotically scale-invariant}. When the system is strictly scale-invariant it is enough to give one element of each row (e.g., $\{r_{N0}\}, \,\forall N$) to fully determine it. If we choose $r_{N0}=1/(N+1)$ we obtain the Leibnitz triangle itself \cite{Polya}, and can verify that it is ``Boltzmannian" in the sense that it is only for $q=1$ that we obtain extensivity. In other words, $S_{BG}(N) \propto N \;(N>>1)$ as can be seen in Fig. 1(a). We may choose a more complex set of probabilities that gives extensivity only for $q=1/2$ as shown in Fig. 1(b). For brevity of space we skip here the details of this set, which is asymptotically scale-invariant. They  details can be seen in \cite{tsallisgellmannsato}. The main reason is, however, quite straightforward. The number of states whose probability differs from zero increases exponentially with $N$ in the former case (Fig. 1(a)), and only as $N^{2}$ in the latter (Fig. 1(b)) 
\footnote{If the number of states whose probability differs from zero increases as $\mu^N$ with $\mu>1$ (as $N^\rho$ with $\rho >0$) the entropy $S_q$ is extensive only for $q=1$ (for $q=1-1/\rho$).}. 
It is then clear that the property $S(A_1+A_2)/k=[S_q(A_1)/k] + [S_q(A_2)/k]+[S_q(A_1)/k][S_q(A_2)/k]$, which led to the term "nonextensive entropy" for $S_q$, is valid {\it only} under the (explicit or tacit) assumption of independence. 

Our overall conclusion at this point is that extensivity does not only depend on the specific functional form of the entropy but {\it also} on the composition law that we are using to form the total system out of its subsystems and their possible correlations. We have exhibited here systems of different nature (and different values of $q$) which, nevertheless, satisfy the Clausius requirement of being extensive. The various sets of probabilities that enter in the calculation of the entropy are, of course, expected to have basically a microscopic nonlinear dynamical origin. Strong chaos (i.e., at least one {\it positive} Lyapunov exponent for classical systems) would typically yield $q=1$, whereas weak chaos (i.e., the maximal Lyapunov exponent vanishes) would typically yield $q \ne 1$, but the analysis of this fundamental point is out of the scope of the present short paper.
\begin{figure} [t]
\includegraphics[width=7.cm,height=7. cm]{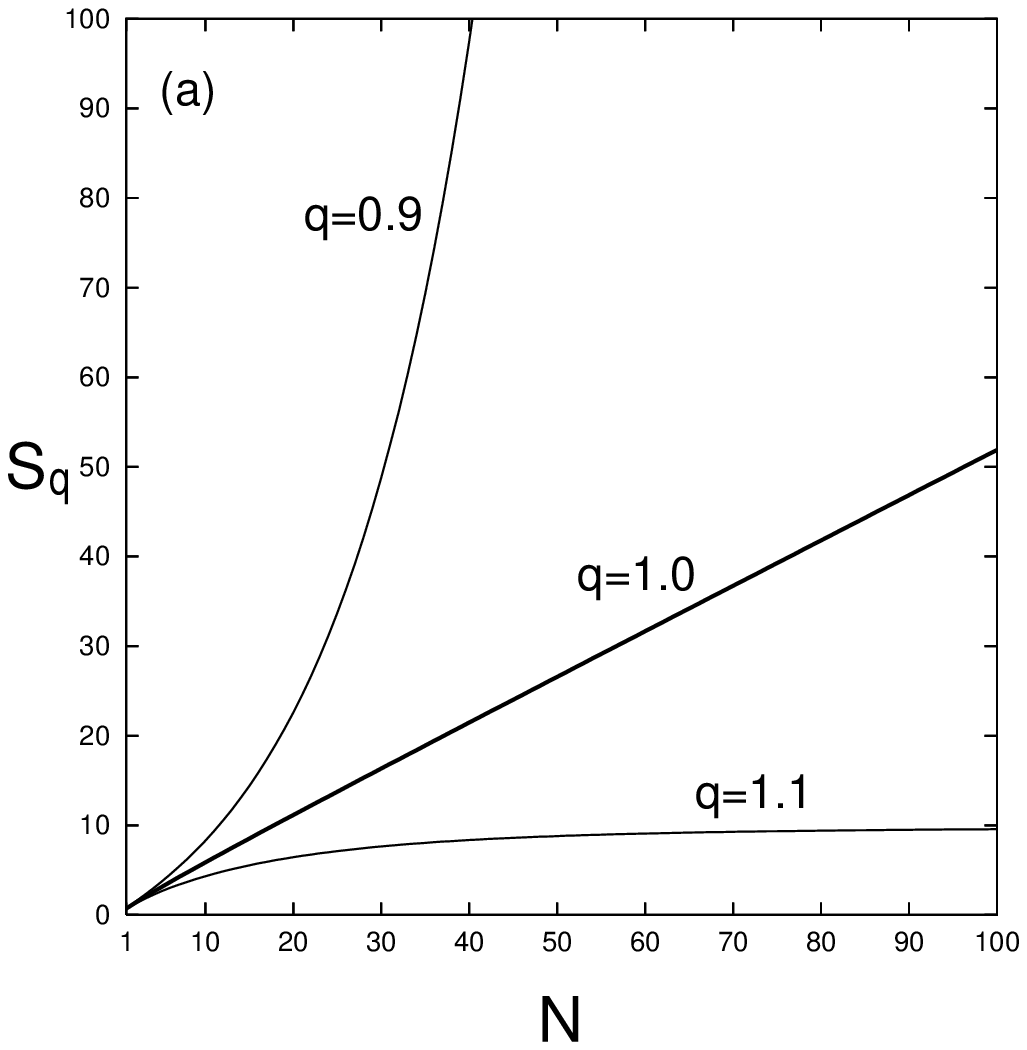}
\includegraphics[width=7.cm,height=7. cm]{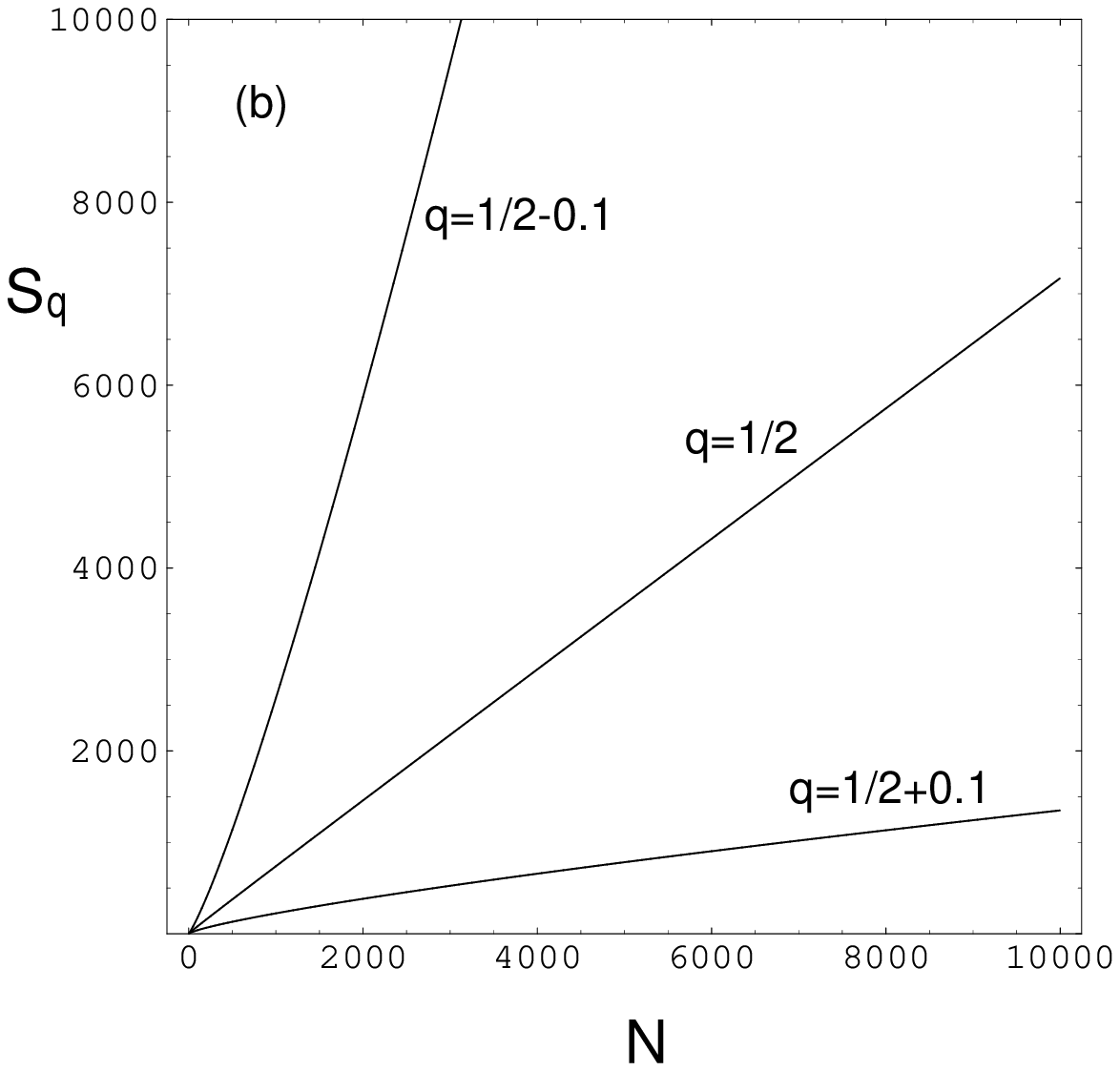}
\caption{Extensivity of $S_q(N)$: (a) only for $q=1$ (Leibnitz triangle); (b) only for $q=1/2$ (complex set of probabilities which is asymptotically scale-invariant). See details in  \cite{tsallisgellmannsato}
}
\end{figure}
\begin{table}[htbp]
\begin{flushleft}
~~~~~~~~~~~~~~~~~~~~~~~~~~~~~~~~~~~~$(N=0)$        ~~~~~~~~~~~~~~~~~~~         ~~~~$1$~~~~\\
~~~~~~~~~~~~~~~~~~~~~~~~~~~~~~~~~~~~$(N=1)$        ~~~~~~~~~~~~~~~~         ~~~$r_{10}$~~$r_{11}$~~~\\ 
~~~~~~~~~~~~~~~~~~~~~~~~~~~~~~~~~~~~$(N=2)$        ~~~~~~~~~~~~~~         ~~$r_{20}$~~$r_{21}$~~$r_{22}$~~\\ 
~~~~~~~~~~~~~~~~~~~~~~~~~~~~~~~~~~~~$(N=3)$        ~~~~~~~~~~~~         ~$r_{30}$~~$r_{31}$~~$r_{32}$~~$r_{33}$~\\ 
~~~~~~~~~~~~~~~~~~~~~~~~~~~~~~~~~~~~$(N=4)$        ~~~~~~~~~~         $r_{40}$~~$r_{41}$~~$r_{42}$~~$r_{43}$~~$r_{44}$\\
\end{flushleft}
\vspace{0.5cm}
\caption{Most general set of joint probabilities for $N$ equal and distinguishable binary subsystems for which only the number of states 1 and of states 2 matters, {\it not their ordering}. These probabilities satisfy $\sum_{n=0}^N \frac{N!}{(N-n) !\,n!}\, r_{Nn}=1$. The particular simple case of {\it independence} corresponds to $r_{Nn}=(\pi_1^{A_1})^{N-n}(1-\pi_1^{A_1})^n$.
} 
\end{table}

\section{On the possible $q$-generalization of the central limit theorem}

In the case of the canonical ensemble we must optimize $S_q$ under the following constraints \cite{TMP} :
\begin{eqnarray}
\sum_{i=1}^W p_i=1,\,\\
\frac{\sum_{i=1}^W E_i p_i^q}{\sum_{i=1}^W p_i^q} = U_q \,,
\end{eqnarray}
where 
$E_i$ is the energy of the $i^{th}$ configuration and $U_q$ some fixed quantity.
The optimizing distribution of energies is given by $p(E_i) = \frac{e_q^{-\beta_q (E_i-U_q)}}{\sum_{j=1}^We_q^{-\beta_q (E_j-U_q)}}$, $\beta_q$ being an inverse temperature-like parameter and $e_q^z \equiv [1+(1-q)\,z]^{1/(1-q)}\;(e_1^z=e^z)$. This expression can be rewritten as $p(E_i) = \frac{e_q^{-\bar{\beta}_q E_i}}{\sum_{j=1}^We_q^{-\bar{\beta}_q E_j}}$, where $\bar{\beta}_q$ is a function of $U_q$. This corresponds to the case when the microscopic configurations are discrete. Its continuous analogous corresponds, in the simplest case, to optimize $S_q=k\frac{1-\int dx\, [p(x)]^q}{q-1}$ with the constraints $\int dx\,p(x)=1$ and  $\frac{\int dx\,E(x) \,[p(x)]^q}{\int dx \,[p(x)]^q}=U_q$. A very elementary case corresponds to $E(x) \propto x^2$, hence $p(x) \propto e_q^{-K \, x^2}$, where $K$ is a positive constant. It happens that a considerable number of natural and artificial cases do exhibit this distribution. This would be neatly understandable if this form is the attractor, in the space of the distributions, under some quite generic type of composition of the random variables that are being averaged, for example if the variables  have a special correlation. If this correlation is scale-invariant in a sense similar to the one that we have discussed in the previous Section, the mathematical structure would be that of a $q$-generalized central limit theorem. The possibility of such a theorem has been addressed in several occasions in the literature \cite{Bologna}, and was specifically analyzed in \cite{milano}. Strong numerical indications became recently available \cite{moyanogellmann} which $q$-generalize the celebrated de Moivre-Laplace theorem. Let us briefly review them. 

We consider a triangle such as that in Table III, and assume the Leibnitz rule to be strictly valid. To determine all the probabilities it is enough, as already mentioned, to provide, for instance, $r_{N0}$. We shall adopt the $q$-product \cite{borges} $r_{N0}= 1/\, [N2^{1-q}-(N-1)]^{1/(1-q)} $ (see details in \cite{moyanogellmann}). It has been verified that, for increasing $N$, the probability distribution approaches, after appropriate centralization, rescaling and symmetrization, a distribution $p(x) \propto e_{q_e}^{-\bar{K}\,x^2}$ with $\bar{K}$ some positive constant and
\begin{equation}
q_e=2-\frac{1}{q} \,\;\;\;(q \le 1).
\end{equation}

Let us introduce the definitions
\begin{eqnarray}
\nu(x) &\equiv& 2-x \,,\\
\mu(x) &\equiv& \frac{1}{x} \,,
\end{eqnarray}
respectively corresponding to {\it additive} and {\it multiplicative dualities} ($\nu^2(x)=\mu^2(x)=x, \,\forall x$). We then see that Eq. (2.3) can be rewritten as $q_e=\nu \mu(q)$. The two transformations $\mu$ and $\nu$ appear (isolated or combined) very frequently (e.g.,  \cite{dualities}) within nonextensive statistical mechanics. They can provide some hint for understanding the meaning of the $q$-triplet that was conjectured some time ago \cite{Tsallistriplet}, and was recently confirmed by Burlaga and Vinas \cite{Burlaga} with data on the solar wind sent by the spacecraft Voyager 1 from the distant heliosphere.

\section{On the $q$-triplet}

The $q$-triplet is defined as $(q_{sen},q_{rel},q_{stat})$ where {\it sen}, {\it rel} and {\it stat} stand for {\it sensitivity} (to the initial conditions), {\it relaxation} and {\it stationary state} respectively (see \cite{Tsallistriplet} for details). Let us assume that we are addressing a classical many-body Hamiltonian system with two-body interactions given by an attractive potential energy which at the origin behaves smoothly and which, at long distances, behaves like $r^{-\alpha}$ ($\alpha >0$). It is reasonable to expect that the $q$-triplet depends on $\alpha$ and on the space-dimension $d$. It is in fact expected that it depends only on the ratio  $\alpha/d$ \cite{integrablerome}. 

Naturally, for $\alpha/d \ge 1$ we expect $q_{sen}=q_{rel}=q_{stat}=1$. But for $0 \le \alpha/d < 1$, the functional forms still are unknown. We expect, however, $q_{sen}(\alpha/d) \le 1 \le q_{stat}(\alpha/d) \le q_{rel}(\alpha/d)$. A simple possibility is as follows:
\begin{eqnarray}
q_{stat}&=&2-\frac{\alpha}{d} \equiv \nu(\alpha/d)    \,,\\
q_{rel}&=&\frac{d}{\alpha}    \;\;\;\;\;\; \equiv \mu(\alpha/d)\,,\\
q_{sen}&=&\frac{\alpha}{d}    \,.
\end{eqnarray} 
This set (see Fig. 2) satisfies the following relations
\begin{eqnarray}
q_{stat}+q_{sen}=2   \,,\\
q_{sen} \,q_{rel}=1    \,,\\
q_{rel}\,(2-q_{stat})=1 \,.
\end{eqnarray} 
Relations such as these might be applicable for Hamiltonian systems.

Another simple possibility is as follows:
\begin{eqnarray}
q_{stat}&=&2-\frac{\alpha}{d} \equiv \nu(\alpha/d)   \,,\\
q_{rel}&=&\frac{d}{\alpha}    \;\;\;\;\;\; \equiv \mu(\alpha/d)  \,,\\
q_{sen}&=&2-\frac{d}{\alpha} \equiv \nu\mu(\alpha/d)    \,.
\end{eqnarray} 
This set (see Fig. 2) satisfies the following relations
\begin{eqnarray}
q_{rel}+q_{sen}=2   \,,\\
q_{rel}\, (2-q_{stat})=1    \,,\\
(2-q_{stat}) \, (2-q_{sen})=1 \,.
\end{eqnarray} 
Relations such as these might be applicable for dissipative systems.

\begin{figure} [t]
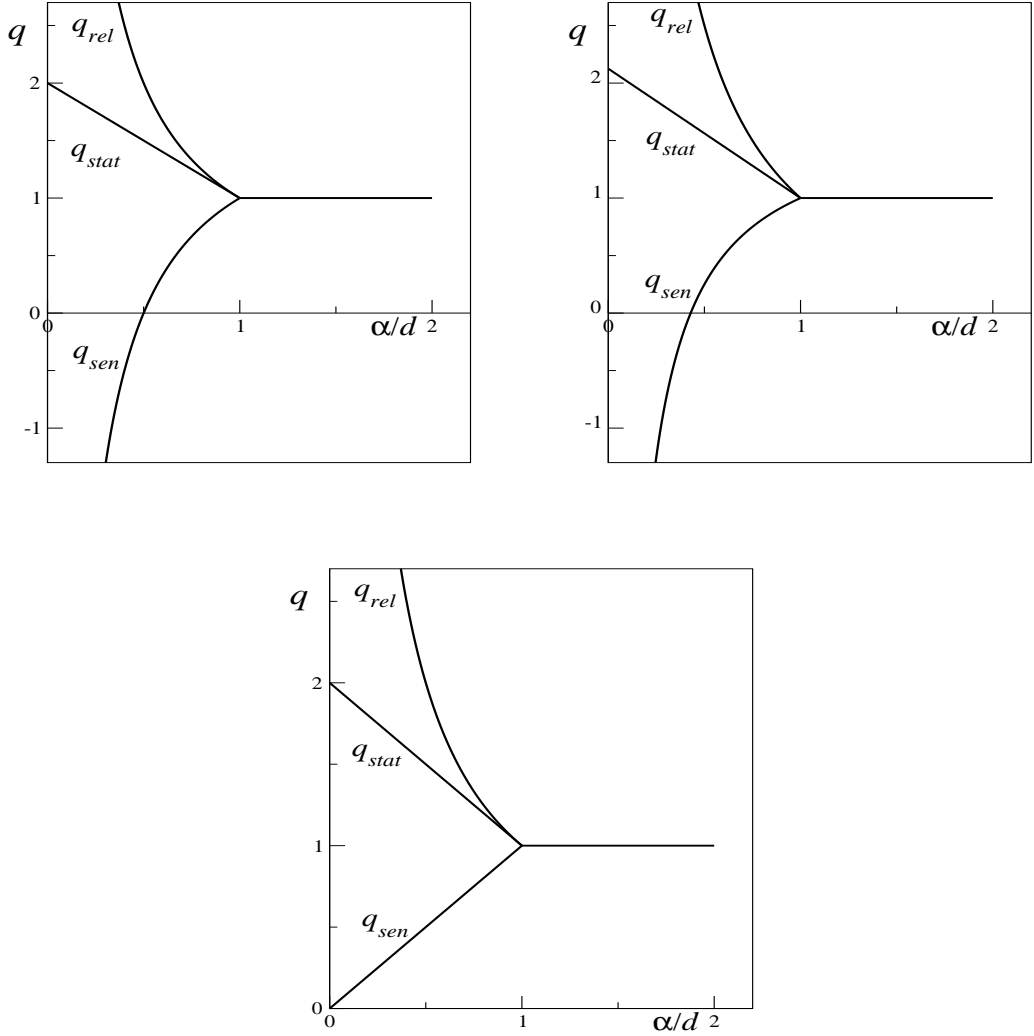

\centerline{
\includegraphics[width=6.2cm,height=6.2 cm]{qtriplet2.eps}
\hspace{1.0 cm}
\includegraphics[width=6.2cm,height=6.2cm]{qtriplet3.eps} \\
\vspace{1.3 cm}
}
\centerline{
\includegraphics[width=6.2cm,height=6.2cm]{qtriplet4.eps}
}
\caption{$\alpha/d$-dependence of the $q$-triplet $(q_{sen},q_{rel},q_{stat})$. {\it Up left:} set of Eqs. (3.7-3.9); {\it Up right:} set of Eqs. (3.16-3.18) {\it Bottom:}set of Eqs. (3.1-3.3).
}
\end{figure}
\begin{figure} [t]
\centerline{
\includegraphics[width=6.2cm,height=6.2 cm]{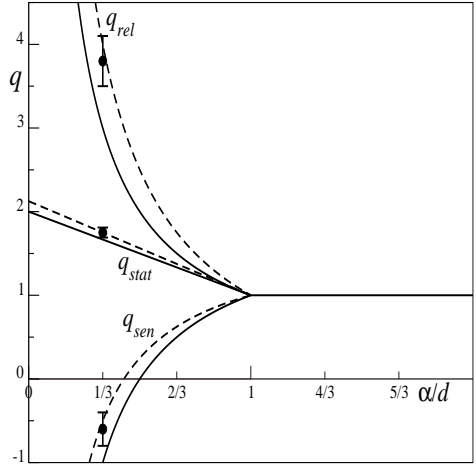}
}
\caption{
Solid and dashed lines correspond to Eqs. (3.7-3.9) and Eqs. (3.16-3.18) respectively. The NASA observational data \cite{Burlaga} have been tentatively located at $\alpha/d=1/3$.
}
\end{figure}
It was determined in \cite{Burlaga} that  $(q_{sen},q_{rel},q_{stat})=(-0.6 \pm 0.2, 3.8 \pm 0.3, 1.75 \pm 0.06)$. Since important dissipative phenomena might be present in the solar wind,
 it seems reasonable to compare the experimental data with conjectures (3.7) to (3.12) (In any case, relation (3.3) is incompatible with the negative value observed for $q_{sen}$.). If we do make such comparison, it appears as physically tempting to use the above possible expressions at their values corresponding to the Coulombian interaction within the solar wind plasma, i.e., $\alpha=1$ and $d=3$, hence $\alpha/d=1/3$.  An acceptably close theoretical set could then be $(q_{sen},q_{rel},q_{stat})=(-1,3,5/3)$. Although outside the observational error bars, we might not wish to exclude it at this early stage of this complex problem. If we wish, however, to get theoretical values closer to the observational ones, we might consider a set of relations such as (3.7-3.9) but somewhat more general in order to have a flexible scenario. We can then analyze the following more general forms:
\begin{eqnarray}
q_{stat}&=& A+(1-A)\frac{\alpha}{d}  \,,\\
q_{rel}&=& B+(1-B)\frac{d}{\alpha}           \,,\\
q_{sen}&=& C +(1-C) \frac{d}{\alpha}  \,.
\end{eqnarray} 
where $A$, $B$ and $C$ are constants. We verify that, by construction, all $q$'s equal unity for $\alpha/d=1$, as desired. For $(A,B,C)=(2,0,2)$ we recover Eqs. (3.7) to (3.9).

It is clear that a set $(A,B,C)$ exists which reproduces the observational data. A different, more simple, possibility is to assume that the triplet is constituted by simple rationals.  In that case a solution is provided by $(q_{sen},q_{rel},q_{stat})=(-1/2,4,7/4)$ which satisfies \cite{tsallisgellmannsato} the following simple relations: $q_{stat}=\nu \mu (q_{rel})$ and $q_{sen}=\mu \nu (q_{rel})$. In this case we have $(A,B,C)=(17/8,-1/2,7/4)$, hence
\begin{eqnarray}
q_{stat}&=&\; \frac{17}{8}- \frac{9}{8}\,\frac{\alpha}{d}  \,,\\
q_{rel}&=& -\frac{1}{2}+\frac{3}{2}\,\frac{d}{\alpha}           \,,\\
q_{sen}&=& \;\;\; \frac{7}{4}-\frac{3}{4}\, \frac{d}{\alpha}   \,.
\end{eqnarray}  
This set (see Fig. 2) satisfies the following relations
\begin{eqnarray}
\frac{q_{rel}+2\,q_{sen}}{3}=1                    \,,\\
\frac{(2\,q_{rel}+1)\,(17-8\,q_{stat})}{27}    =1     \,,\\
\frac{(7-4\,q_{sen})\,(17-8\,q_{stat})}{27}=1                      \,.
\end{eqnarray} 

The sets (3.7-3.9) and (3.16-3.18) are numerically not very different: see Fig. 3. The former is mathematically extremely simple; the latter coincides with the experimental data within the error bars. We have not succeeded at this point to find further arguments for better understanding this complex phenomenon. Consistently, this entire Section cannot be considered as more than a very rough first discussion of the issue.

\section{Conclusions}

We have illustrated that the extensivity of $S_q$ for $q \ne 1$ is possible in the presence of relevant correlations (e.g., strictly or asymptotically scale invariant ones) between the elements of the system. This seems to be consistent with the existence of a $q$-generalized central limit theorem, in the same way the standard, Gaussian, central limit theorem plays a crucial role in Boltzmann-Gibbs statistical mechanics. Some numerical indications already exist that point towards such theorem. These indications provided a very simple relation based on additive and multiplicative dualities. This in turn enabled a first exploratory analysis of the values determined, for the $q$-triplet, from the data obtained for the solar wind at the distant heliosphere. 

One of the many interesting problems which remain open at this stage is the illustration of the $q$-generalized central limit theorem with $q_e>1$. A $q$-generalization of the de Moivre-Laplace theorem that would lead to say $q_e=3-2/q$ ($q \ge 1$) would certainly be very interesting. Indeed, such a relation implies that $q$ increasing from unity to infinity makes $q_e$ to increase from unity to 3, the upper limit for $q_e$-Gaussians to be normalizable.  Also, $q \ge 3/2$ implies $q_e \ge 5/3$, which corresponds to diverging variance of the $q_e$-Gaussians.

\section*{Acknowledgements}
Warm hospitality by S. Abe in Japan is heartily acknowledged. I have benefited  from long  discussions with M. Gell-Mann on virtually all the points presented here. Useful remarks from E.G.D. Cohen, J.D. Farmer, F. Lillo, L.G. Moyano and Y. Sato  are acknowledged as well. Sponsoring from SI International and Air Force Research Laboratory is also acknowledged.

\end{document}